\documentclass[12pt]{article}
\usepackage{authblk}
\usepackage[numbers]{natbib}
\usepackage{times}
\usepackage{float}
\usepackage{amsmath}
\usepackage{graphicx}
\usepackage{epsfig}
\usepackage{lineno}
\usepackage{url}
\usepackage[breaklinks,colorlinks,citecolor=blue]{hyperref}
\usepackage{cleveref}
\usepackage[all]{hypcap}
\usepackage[vmargin={1in,1in},hmargin={1in,1in}]{geometry}
\usepackage{setspace}

\begin{document}

\title{A Large Repository of 3D Climate Model Outputs for Community Analysis and Postprocessing}
\author[1,*]{Adiv Paradise}
\author[1,2,3]{Bo Lin Fan}
\author[2]{Evelyn Macdonald}
\author[1,2,4]{Kristen Menou} 
\author[2]{Christopher Lee} 

\affil[1]{David A. Dunlap Department of Astronomy and Astrophysics, University of Toronto, St. George, Toronto, Ontario, Canada}
\affil[2]{Department of Physics, University of Toronto, St. George, Toronto, Ontario, Canada}
\affil[3]{Department of Ecology \& Evolutionary Biology, University of Toronto, St. George, Toronto, Ontario, Canada}
\affil[4]{Department of Physical and Environmental Sciences, University of Toronto, Scarborough, Toronto, Ontario, Canada}

\affil[*]{Corresponding author email: paradise@astro.utoronto.ca}

\maketitle
%
%
%

As the number of known exoplanets has climbed into the thousands, efforts by theorists to understand the diversity of climates that may exist on terrestrial planets in the habitable zone have also accelerated. These efforts have ranged from analytical, to simple 0-D, 1-D, and 2-D models, to highly-sophisticated 3D global climate models (GCMs) adapted from Earth climate and weather models. The advantage of the latter is that fewer physical processes are reduced to simple parameterizations and empirical fits, and may instead be represented by physically-motivated algorithms. However, many such models are difficult to use, and take a long time to reach a converged state relative to simpler models, thereby limiting the amount of parameter space that can be explored. We use PlaSim, a 3D climate model of intermediate complexity, to bridge this gap, allowing us to produce hundreds to thousands of model outputs that have reached energy balance equilibrium at the surface and top of the atmosphere. We are making our model outputs available to the community in a permanent Dataverse repository (\href{https://dataverse.scholarsportal.info/dataverse/kmenou}{https://dataverse.scholarsportal.info/dataverse/kmenou}). A subset of our model outputs can be used directly with external spectral postprocessing tools, and we have used them with petitRADTRANS and SBDART in order to create synthetic observables representative of fully-3D climates. Another natural use of this repository will be to use more-sophisticated GCMs to cross-check and verify PlaSim's results, and to explore in more detail those regions of the exoplanet parameter space identified in our PlaSim results as being of particular interest. We will continue to add models to this repository in the future, including more than 1000 models in the short- to medium-term future, expanding the diversity of climates represented therein.

PlaSim \citep{Fraedrich2005} uses a spectral core to solve the primitive equations in the atmosphere, and couples the atmosphere to a mixed-layer slab ocean through precipitation, evaporation, and latent heat fluxes. A sea ice model is included that allows for changes in ice thickness, snowfall on top of ice, and thermodynamic changes resulting from interactions between snow and seawater. A soil model is also included for land areas, which uses buckets to hold groundwater, and advects excess water toward continental margins to represent rivers. Snow can also accumulate on land, and contributes either to groundwater or evaporation when it melts. Radiation in PlaSim is represented by two shortwave bands ($<$0.75$\mu$m, SW1, and $>$0.75$\mu$m, SW2) and one longwave band. Rayleigh scattering is confined to the SW1 band, and shortwave absorption by water vapor is confined to SW2. Near-infrared and longwave absorbers are treated as gray absorbers and emitters, with parameterized absorptivities and emissivities. The total absorption, emission, and scattering therefore depends on the abundance of the different species, and while CO$_2$ is by default treated as well-mixed, water vapor is advected through tracer transport. Ozone is prescribed as having a Gaussian abundance in the stratosphere that varies with altitude, and as PlaSim does not natively model much of the stratosphere, its effects are simply represented by absorption in the upper layers of the model. 

PlaSim's default horizontal resolution is T21, or 32 latitudes and 64 longitudes (roughly 5.6$^\circ$ in each direction). In most of our models, unless specified otherwise, we use 10 vertical levels. PlaSim also supports T42 resolution (64 latitudes and 128 longitudes), and we have had some success in running at even higher resolutions. The model has been used extensively to study Earth-like climates, snowball climates, Martian analogues, and tidally-locked planets \citep[e.g.][]{Boschi2013,Checlair2017,Fraedrich2005,Nowajewski2018,Paradise2017,Paradise2019,segschneider2005}. We have modified the model to expand the range of planets that can be modeled to include planets with surface pressures significantly different from Earth's \citep{ParadiseN2,exoplasim}. The modified model is available at AP's github at \href{https://github.com/alphaparrot/ExoPlaSim}{https://github.com/alphaparrot/ExoPlaSim} \citep{exoplasim}. In our computing environment, we find that, at T21 and with 10 vertical layers, PlaSim can simulate one year of climate on 16 CPUs in 30-60 seconds, allowing models to run for a few centuries to energy-balance equilibrium in a matter of hours, to days at most.

\begin{figure}
\begin{center}
\includegraphics[width=6.5in]{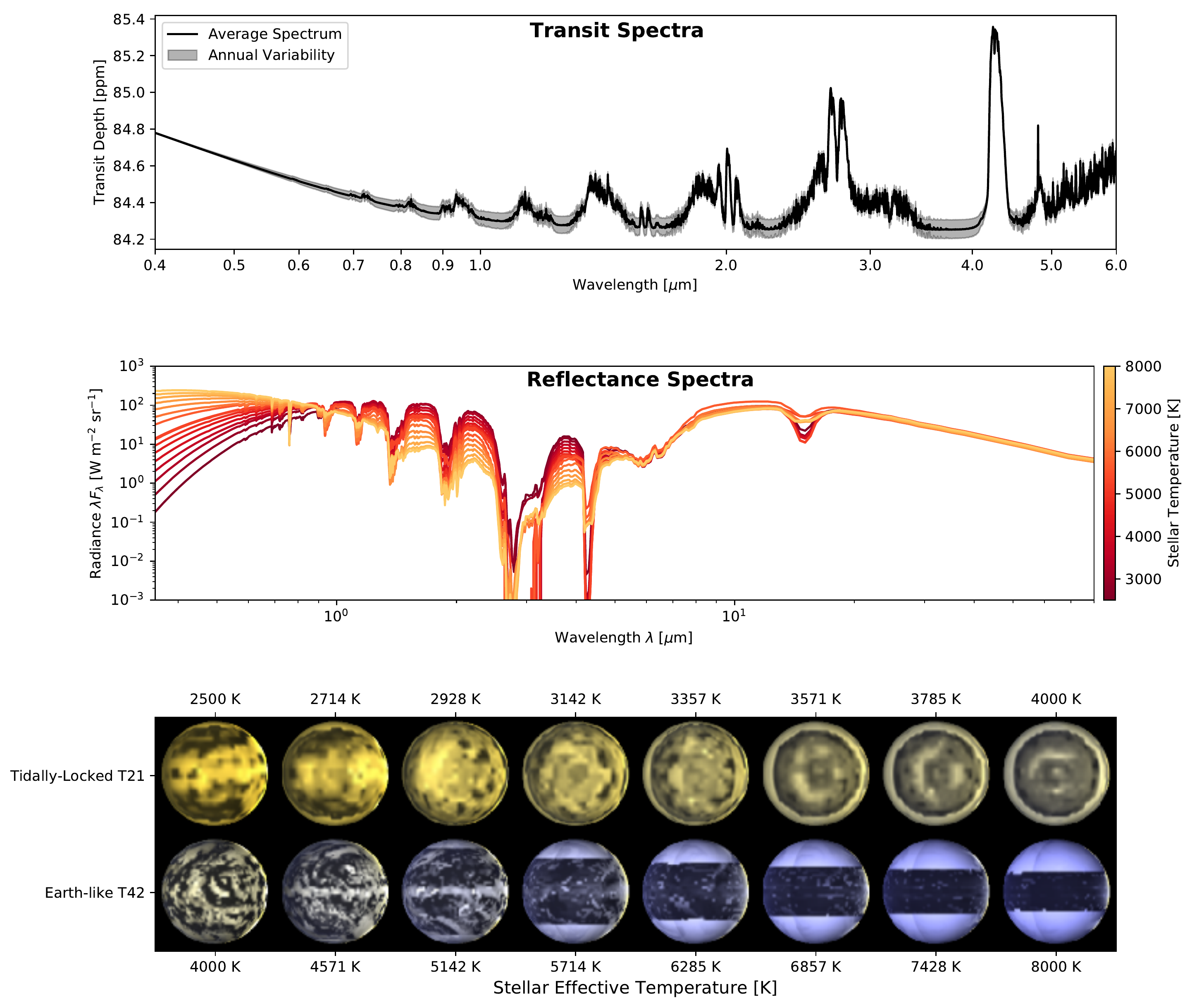}
\end{center}
\caption{Examples of observables that can be computed using PlaSim snapshot outputs. The top panel depicts transit spectra for an Earth-sized, Earth-rotating aquaplanet computed using petitRADTRANS, with both the annual mean and variability range shown. The middle and bottom panels were computed with SBDART, using a set of aquaplanets orbiting stars ranging from 2500 K effective temperature to 8000 K, with the planets at 4000 K and below being tidally-locked and run at 5.6$^\circ$ resolution, and those at 4000 K and above having 24-hour rotation and zero obliquity, and run at 2.8$^\circ$ resolution. The middle panel represents disk-integrated spectra, and the bottom panel shows orthographic projections of the individual spectra transformed into an RGB colorspace. Each curve in the middle panel is the disk-integrated reflection spectrum of one of the planets shown in the bottom panel, with the color chosen to indicate the effective temperature of the parent star.}\label{fig:sample}
\end{figure}

At present we have output files representing 760 models present in the repository, in a dataset associated with \citet{ParadiseN2}, in which we systematically explored the response of the climate to different N$_2$ partial pressures across a range of insolations. Output files may either include monthly averages of model variables, or snapshots in time which may be used to compute synthetic observables and compare the performance of other radiation codes relative to PlaSim's own. Output files are in the NetCDF format, and can be accessed in Python via the \texttt{netCDF4} package. Output files contain humidity, temperature, wind speed, divergence, vorticity, radiative and thermodynamic fluxes, surface parameters (surface type, snow depth, ice thickness, groundwater depth, albedo, etc), and a number of additional derived quantities. A brief guide to opening and plotting PlaSim output is available as \href{https://github.com/alphaparrot/ExoPlaSim/blob/master/plasim/doc/gcm_analysis.pdf}{part of the ExoPlaSim package} on AP's github.

The model outputs that represent snapshots in time can be used to create boundary conditions and inputs for spectral postprocessing packages, such as petitRADTRANS \citep{Molliere2019} and SBDART \citep{Ricchiazzi1998}. We have used these two packages in our work, and examples of the spectra and results that can be produced are shown in \autoref{fig:sample}. While we have used and tested these two packages with PlaSim outputs, other publicly-available packages and retrieval frameworks should work as well, and the use of multiple packages may serve as a useful point of comparison. The 3D, time-resolved nature of the output means that transit spectra can be computed independently along different point of the terminator and then combined to create the planet's overall transit spectrum, allowing for the realistic inclusion of local and temporal variability such as clouds and variations in jet streams. Similarly, reflected-light spectra that make use of every column on the observer-facing hemisphere allow for the study of partly-cloudy atmospheres, where the outcome is a combination of clear-sky spectra that probe the planet's surface and cloudy spectra that do not. These spatially-resolved spectra can also be integrated and convolved into RGB colorspaces through the use of color-matching functions, allowing for the creation of true-color images that may be useful for diagnosing problems in model output or in one's postprocessing pipeline, as well as aiding in public outreach efforts. The volume of parameter space we sample with PlaSim means that studies that focus on postprocessing efforts will be able to generate large libraries of synthetic observables, aiding in statistical approaches to the search for habitable planets \citep{Checlair2019}.

In addition to enabling easier study of the observational features of Earth-like planets, we also envision our repository aiding in theoretical studies using slower but more-sophisticated GCMs. Such models are computationally much more expensive than PlaSim, and therefore the choice of boundary conditions and configuration parameters is much more important. By off-loading the work of surveying parameter space to PlaSim, teams working with larger GCMs are therefore able to focus their computational efforts on regions of parameter space most likely to probe the phenomena of interest, or which are most likely to yield the climate outcomes relevant to their hypotheses. At the same time, studies which sparsely re-sample the PlaSim-surveyed parameter space with higher-complexity GCMs will be useful in verifying PlaSim's results, as well as in helping to indicate where PlaSim is inaccurate or likely has missing physics (such as a dynamic ocean or sea ice drift). 

It is therefore our belief that PlaSim's ability to both model a range of climates at an intermediate complexity level and run fast enough to produce tens of thousands of climate-years per week of user time makes it a particularly valuable resource for efforts within the community to uncover qualitative climate trends within the exoplanet parameter space and to explore statistical trends in potential observables. It is our hope that by providing a repository which will be periodically updated with additional datasets containing yet more models produced in the course of our own research, we can help to accelerate these efforts.


\end{document}